\documentclass[preprintnumbers,amsmath,amssymb,nofootinbib,superscriptaddress,floatfix,showkeys,twocolumn]{revtex4-1}
\usepackage{graphicx,color}
\usepackage[latin1]{inputenc}
\usepackage{amsmath,amssymb}
\usepackage{hyperref}
\usepackage{epstopdf}
\usepackage{slashed}
\usepackage{soul}
\usepackage{amssymb}
\usepackage[normalem]{ulem}
\usepackage[caption=false]{subfig}

\newcommand{\lsim}{\mathrel{\mathop{\kern 0pt \rlap
  {\raise.2ex\hbox{$<$}}}
  \lower.9ex\hbox{\kern-.190em $\sim$}}}
\newcommand{\gsim}{\mathrel{\mathop{\kern 0pt \rlap
  {\raise.2ex\hbox{$>$}}}
  \lower.9ex\hbox{\kern-.190em $\sim$}}}

\newcommand{\sigmav}{\ensuremath{\langle\sigma v\rangle}}

\newcommand{\sigsip}{\ensuremath{\sigma^{\rm{SI}}_p}}

\newcommand{\gev}{\ensuremath{\,\mathrm{GeV}}}
\newcommand{\tev}{\ensuremath{\,\mathrm{TeV}}}

\newcommand{\Sm}{\ensuremath{\mathcal{S}}}
\newcommand{\Tm}{\ensuremath{\mathcal{T}}}
\newcommand{\Um}{\ensuremath{\mathcal{U}}}

\def  \bcen   {\begin{center}}
\def  \ecen   {\end{center}}
\def  \beq    {\begin{equation}}
\def  \eeq    {\end{equation}}
\def  \beqa   {\begin{eqnarray}}
\def  \eeqa   {\end{eqnarray}}

\def\bea{\begin{eqnarray}}
\def\eea{\end{eqnarray}}

\begin{document}

\title{Inert Higgs Dark Matter for CDF-II $W$-boson Mass and Detection Prospects}
\author{Yi-Zhong Fan}
\author{Tian-Peng Tang}
\author{Yue-Lin Sming Tsai}
\affiliation{Key Laboratory of Dark Matter and Space Astronomy, Purple Mountain Observatory, Chinese Academy of Sciences, Nanjing 210033, China}
\affiliation{School of Astronomy and Space Science, University of Science and Technology of China, Hefei, Anhui 230026, China}

\author{Lei Wu}
\affiliation{Department of Physics and Institute of Theoretical Physics, 
Nanjing Normal University, Nanjing, 210023, China }
\date{\today}

\begin{abstract}
The $W$-boson mass, which was recently measured at FermiLab with an unprecedented precision, suggests the presence of new multiplets beyond the Standard Model (SM).
One of the minimal extensions of the SM is to introduce an additional scalar doublet in which the non-SM scalars can enhance $W$-boson mass via the loop corrections. 
On the other hand, with a proper  discrete symmetry, 
the lightest new scalar in the doublet can be stable and play the role of a dark matter particle.
We show that the inert two Higgs doublet model can naturally handle the new $W$-boson mass without violating other constraints and that the preferred dark matter mass is between $54$ and $74$ GeV. We identify three feasible parameter regions for the thermal relic density: the $SA$ coannihilation, the Higgs resonance, and the $SS \to WW^*$ annihilation. 
We find that the first region can be fully tested by the High Luminosity Large Hadron Collider,
the second region will be tightly constrained by direct detection experiments, and the third region could yield detectable GeV $\gamma$-ray and antiproton signals in the Galaxy 
that may have been observed by Fermi Large Area Telescope and the Alpha Magnetic Spectrometer AMS-02 experiment.
\end{abstract}

\maketitle

%%#######################################################%%
\section{Introduction \label{sec:intro}}
%%#######################################################%%

%{\it Introduction.} 
The origin of mass is one of the most fundamental problems in modern physics. The central idea of generating masses of the electroweak gauge bosons in the standard model (SM) is the spontaneous symmetry breaking (SSB). Therefore, the precision measurement of the gauge boson masses is of great importance in testing the SSB mechanism. With the full dataset, the Collider Detector at Fermilab (CDF) collaboration has recently reported their newly measured $W$-boson mass
$m_{W, {\rm CDF-II}}= 80.4335 \pm 0.0094 {\rm GeV}$~\cite{CDF:2022hxs}, which deviates from the SM prediction $m_{W, {\rm SM}}=80.357\pm 0.006\gev$~\cite{ParticleDataGroup:2020ssz} about $7\sigma$. 
This new measurement, characterized by its  unprecedented precision, slightly conflicts with some previous measurements~\cite{ALEPH:2013dgf,LHCb:2021bjt,ATLAS:2017rzl,D0:2012kms}. If confirmed in the future, as assumed in this Letter, the CDF II $W$-boson mass excess would strongly indicate the presence of new physics related with the SSB, such as models with extended Higgs sectors.

On the other hand, the existence of dark matter (DM) has been favored by various astrophysical and cosmological observations. For decades, Weakly Interacting Massive Particles (WIMPs) are considered as the strongest candidate for DM. One of the minimal scalar WIMP dark matter models is the inert two Higgs doublet model (i2HDM)~\cite{Deshpande:1977rw,Ma:2006km, Barbieri:2006dq, LopezHonorez:2006gr}, where one doublet $H_1$ is the SM Higgs doublet and the other doublet $H_2$ is hidden in the dark sector, given by 
\beq
H_1 =
   \left( \begin{array}{c}  G^+ \\ 
   \frac{1}{\sqrt{2}} \left( v +
h + i G^0 \right)  \end{array}
     \right),~~
H_2 =    \left( \begin{array}{c}  H^+ \\ 
\frac{1}{\sqrt 2} (S  + i A)  \end{array}  \right). 
\nonumber\\
\eeq
Here, $G^\pm$ and $G^0$ are the charged and neutral Goldstone bosons, and $v\approx 246\gev$ is the vacuum expectation value of the SM Higgs field.  
The discrete $\mathbb{Z}_2$ symmetry ($H_1\rightarrow H_1$ and $H_2\rightarrow -H_2$) is introduced to ensure the lightest scalar stable and cannot be spontaneously broken, i.e., $\langle H_2\rangle=0$. 
After the SSB, there will be five physical mass eigenstates, including two $CP$-even Higgs bosons $h$ and $S$, one $CP$-odd Higgs boson $A$, and a pair of charged Higgs $H^{\pm}$. The $\mathbb{Z}_2$-even scalar $h$ is identified as the SM Higgs boson, and $\mathbb{Z}_2$-odd scalar $S$ or $A$ can be the DM particle. Throughout this Letter, we assume that $S$ is lighter and hence the DM particle. Because of the symmetry of exchanging $S$ and $A$, 
the results will be unchanged for $A$ being the 
DM~\cite{Arhrib:2013ela,Goudelis:2013uca,Ilnicka:2015jba,Diaz:2015pyv,Belyaev:2016lok,Tsai:2019eqi,Dolle:2009fn,Banerjee:2021anv,Banerjee:2021hal,LopezHonorez:2010tb,Banerjee:2021xdp}.

In the i2HDM, the origin of the $W$-boson mass is still the Higgs mechanism, though due to contributions from the new Higgs bosons and interactions. Our model predicts deviations from the SM $W$-boson mass that can match that measured by CDF II. The corrections of the non-SM Higgs bosons to the squared $W$-boson mass can be expressed in term of the oblique parameters $\Sm$, $\Tm$, and $\Um$~\cite{Peskin:1991sw,Eriksson:2009ws}, i.e.,
\begin{eqnarray}
\Delta m_W^2=\frac{\alpha c_W^2 m_Z^2}{c_W^2-s_W^2}
\left[-\frac{\Sm}{2}+c_W^2 \Tm +\frac{c_W^2-s_W^2}{4 s_W^2}\Um\right], 
\label{eq:mW}
\end{eqnarray}
where $c_W$ and $s_W$ are cosine and sine of Weinberg angle. 
The fine structure constant and $Z$ boson mass are denoted as $\alpha$ and $m_Z$. 
In Eq.~\eqref{eq:mW}, the dominant contribution to the $W$-boson mass arises from $\Tm$ parameter, 
which is sensitive to the mass splitting of new particles running in the loop of gauge boson self-energy. 
%\yzf{A $m_A/m_{H}^{\pm}\leq 1$ is needed to boost $\Delta m_W^2$.}
Meanwhile, when the charged Higgs bosons are lighter than the neutral Higgs bosons, 
they produce a positive $\Sm$ and thus reduce the corrections to $W$-boson mass. 
Therefore, a heavy $H^\pm$ but light $S$ and $A$ are preferred to enhance the value of $\Delta m_{W}^{2}$. 
Indeed, we will show that the i2HDM with $m_S/m_{H^\pm}<0.5$ and $m_A/m_{H^\pm}<1$ can naturally account for 
the $W$-boson mass anomaly and offer a successful thermal WIMP paradigm without violating other constraints. 
The upcoming collider and DM experiments will be able to effectively probe the favored regions.

\section{Methodology}
%{\it Methodology.}
The scalar potential of i2HDM can be written as
\begin{eqnarray}
\label{eq:potential}
V &=& \mu_1^2 |H_1|^2 + \mu_2^2 |H_2|^2 + \lambda_1 |H_1|^4
+ \lambda_2 |H_2|^4  \nonumber \\
&& +  \lambda_3 |H_1|^2 |H_2|^2  + \lambda_4 |H_1^\dagger H_2|^2 \nonumber \\
&& +\frac{\lambda_5}{2} \left\{ (H_1^\dagger H_2)^2 + {\rm h.c.} \right\} \;\; .
\end{eqnarray}
Taking  $m_h=125\gev$ and  $v=246\gev$,  
there are six free parameters in the scalar potential after the electroweak symmetry breaking, 
namely $\mu_2^2$, $\lambda_2$, $\lambda_3$, $\lambda_4$ and $\lambda_5$.  
Note that $\lambda_2$ is a phenomenologically invisible interaction at the tree-level, 
which is only involved in the four-points interaction of $\mathbb{Z}_2$-odd scalar bosons. 
The relationships between the other four parameters and the physical masses are given by 
\begin{eqnarray}
\label{mh-lambda1}
&& m_h^2=-2 \mu_1^2=2 \lambda_1 v^2, \nonumber\\
&& m_S^2=\mu_2^2+\frac{1}{2} (\lambda_3 +\lambda_4 + \lambda_5) v^2=
\mu_2^2+ \lambda_S v^2, \nonumber\\
&& m_A^2=\mu_2^2+\frac{1}{2} (\lambda_3 +\lambda_4 - \lambda_5) v^2 =
\mu_2^2+ \lambda_A v^2, \nonumber\\
&& m_{H^\pm}^2= \mu_2^2+ \frac{1}{2} \lambda_3  v^2,  
\end{eqnarray}
where $\lambda_S$ and $\lambda_A$ represent the $hSS$ and $hAA$ couplings, respectively. For convenience, we use the mass splitting parameters $\Delta^0 =m_A -m_S$ and
$\Delta^{\pm} =m_{H^{\pm}} -m_S$ 
to study the new contributions to the $W$-boson mass and DM relic density in the i2HDM. In the following investigation, we choose the input parameters as $\{m_S,\Delta^0, \Delta^{\pm}, \lambda_2 , \lambda_S \}$.

We explore the parameter space of i2HDM with the Markov chain Monte Carlo method in the ranges of
\begin{eqnarray}
30.0 &\leq \, m_S  / \textrm{ GeV} \, \leq & 4000.0, \nonumber\\
10^{-4} &\leq \, \Delta^0  / \textrm{ GeV} \, \leq & 500.0, \nonumber\\
1.0 &\leq \, \Delta^{\pm}  / \textrm{ GeV} \, \leq & 500.0,\nonumber\\
-1.0 &\leq \, \lambda_S  \, \,\leq& 1.0, \nonumber\\
10^{-10} &\leq \, \lambda_2  \,\leq& 4.2. 
\label{eq:domain}
\end{eqnarray}
We calculate the mass spectrum, theoretical bounds on the Higgs potential, and electroweak precision observables with \texttt{2HDMC}~\cite{Eriksson:2009ws}. Since the observed DM relic density and DM direct detection provide the stringent constraints, we compute the DM observables such as the relic density, the annihilation cross section, and 
the spin-independent DM-nucleon scattering cross section with \texttt{micrOMEGAs}~\cite{Belanger:2018mqt}. We also consider the collider constraints from the null results of searching for new scalar bosons, exotic Higgs decays, mono-X searches, and Higgs decay to diphoton $R_{\gamma\gamma}$ as in Ref.~\cite{Tsai:2019eqi}. 

%******************************************************************
\begin{table}[t!]
    \centering
    \begin{tabular}{|l|l|}
    \hline\hline
        Likelihood type & Constraints \\
        \hline
        Step &  perturbativity, stability, unitarity~\cite{Eriksson:2009ws}\\
        Step & LEP-II~\cite{Agashe:2014kda}, OPAL~\cite{Abbiendi:2003ji}\\
        Half-Gaussian & PandaX-4T~\cite{PandaX-4T:2021bab}\\
        Half-Gaussian &  exotic Higgs decays~\cite{Aaboud:2019rtt}\\
        Gaussian & relic abundance~\cite{Aghanim:2018eyx}\\
        Gaussian & $R_{\gamma\gamma}$~\cite{ATLAS:2018doi}\\
        \hline
        Gaussian & EWPT~\cite{ParticleDataGroup:2020ssz}  or  $m_{W,{\rm CDF-II}}$~\cite{CDF:2022hxs}\\   
        \hline\hline
    \end{tabular}
    \caption{Likelihood distributions and constraints used in our analysis.}
    \label{tab:likelihood}
\end{table}
%*******************************************************************

In order to present the allowed parameter space, we use ``\textit{Profile Likelihood}'' method~\cite{Rolke:2004mj} 
to get rid of nuisance parameters while showing the two dimensional contours.
In Table~\ref{tab:likelihood}, we list the above experimental constraints incorporated in our likelihood functions. 
The total $\chi^2_\mathrm{tot}$ is to sum over the individual $\chi^2$ of these constraints. 
We use Gaussian likelihood with 
\begin{equation}
    \chi^2 = \left( \frac{\mu - \mu_\mathrm{exp}}{\sigma} \right)^2~{\rm and}~
    \sigma = \sqrt{\sigma^2_\mathrm{theo} + \sigma^2_\mathrm{exp}},
\end{equation}
where $\mu$ is the theoretical prediction and $\mu_\mathrm{exp}$ is the experimental central value.  
The uncertainty $\sigma$ includes both theoretical and experimental errors. For those Half-Gaussian functions, we can set $\mu_\mathrm{exp}=0$ based on the null signal. We use the hard cuts for the theoretical bounds, Large Electron-Position collider (LEP-II), and Omni-Purpose Apparatus at LEP (OPAL) limits.

To examine the impact of the new CDF II $m_W$ measurement, 
we perform two sets of numerical scans by taking two different likelihoods for electroweak precision data. 
Please bear in mind that these two scans share the same constraints in Table~\ref{tab:likelihood},  
except for electroweak precision likelihood.
The first likelihood, which is denoted as PDG2020 and does not take into account the latest CDF-II $m_W$ data, 
includes the previous complete electroweak precision measurements 
that are parameterized by three oblique parameters $\Sm=-0.01\pm 0.1$, $\Tm=0.03\pm 0.12$ and 
$\Um=0.02\pm 0.11$ (all intervals are for 68\% confidence level). 
The correlation coefficients of $(\Sm, \Tm)$, $(\Sm,\Um)$ and $(\Tm,\Um)$ are $0.92$, $-0.8$, and $-0.93$, respectively~\cite{ParticleDataGroup:2020ssz}.
We refer the readers to Ref.~\cite{Tsai:2019eqi} for the implementation of the covariance matrix with oblique parameters. 
For the second, 
we use the CDF II $W$-boson mass measurement $m_{W, {\rm CDF-II}}=80.4335\pm 0.009\gev$ 
as the electroweak precision test likelihood function.

On the other hand, since the DM indirect detection constraints likely suffer from some systematic uncertainties, we will not include them in the likelihood but compare our allowed parameter space with the limits set by the Fermi Large Area Telescope observations of dwarf Spheroidal galaxies~\cite{Fermi-LAT:2016uux} as well as the signal regions of the Fermi Large Area Telescope Galactic Center $\gamma$-ray excess~\cite{Hooper:2010mq,Zhou:2014lva,Calore:2014xka,Daylan:2014rsa}, and the Alpha Magnetic Spectrometer AMS-02 experiment antiproton excess~\cite{1610.03840,1610.03071,1803.02163,1903.02549}. 

We adopt the Markov chain Monte Carlo scans by using the code \texttt{EMCEE}~\cite{ForemanMackey:2012ig}. To reach a good coverage of the parameter space, we perform several scans and finally collect $\mathcal{O}(4.5\times 10^6)$ data points. The confidence intervals are calculated from the tabulated values of $\Delta\chi^2\equiv -2\ln(\mathcal{L/L}_{\rm{max}})$. For a two-dimensional plot, the 95\% confidence ($2\sigma$) region is defined by $\Delta\chi^2 \leq 5.99$ under the assumption of approximate Gaussian likelihood.

%%%%%%%%%%%%%%%%%%%%%%%%%%%%%%%%%%%%
\section{Numerical Results and Discussions}
%\label{sec:result}
%%%%%%%%%%%%%%%%%%%%%%%%%%%%%%%%%%%%

%{\it Numerical Results and Discussions.} 
As mentioned above, we present two sets of results based on the likelihoods summarized in Table~\ref{tab:likelihood}. One is in gray (see Fig.1 and 2), which is obtained from the global fit with PDG2020 electroweak precision test (EWPT)~\cite{ParticleDataGroup:2020ssz}. While the other, marked in green, blue and red, takes into account the new CDF II $m_W$ data in the fit~\cite{CDF:2022hxs}.

\begin{figure}
\includegraphics[width=8cm,height=8cm]{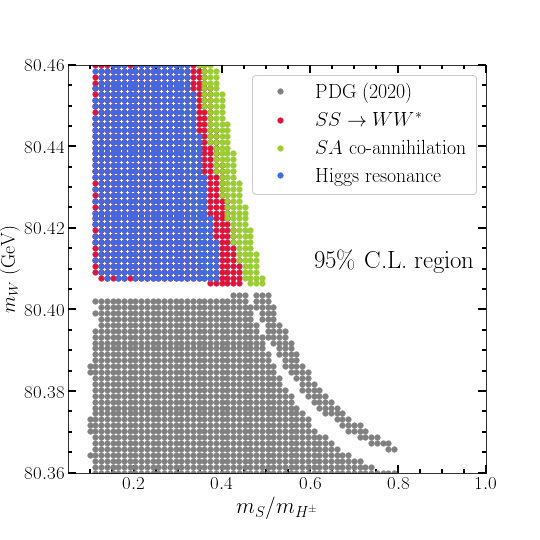}
\includegraphics[width=8cm,height=8cm]{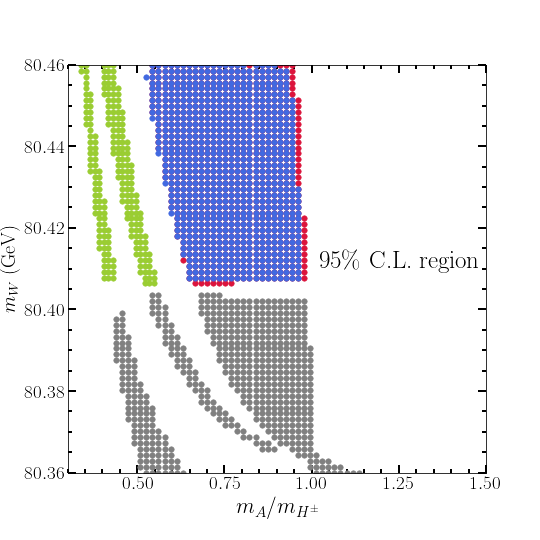}
\caption{The 95\% confidence level allowed regions for three DM production mechanisms: $SA$ coannihilation (green), the Higgs resonance (blue), and $S S\to W W^*$ annihilation (red) on the plane of $m_{S}/m_{H^\pm}$ versus $m_{W}$, and $m_{A}/m_{H^\pm}$ versus $m_{W}$. 
The gray regions are allowed by the data of PDG2020 but excluded by $m_{W,{\rm CDF-II}}$. 
The $95\%$ region of $m_{S}/m_{H^\pm}$ has been narrowed down to $\leq 0.5$.}
\label{fig:mass_ratio}
\end{figure}

In Fig.~\ref{fig:mass_ratio}, we display the $95\%$ allowed regions on the planes of $m_{S}/m_{H^\pm}$ versus $m_{W}$ 
as well as $m_{A}/m_{H^\pm}$ versus $m_{W}$ with and without constraint of the latest CDF II $m_W$ measurement. 
The color regions present the favored dominant DM production mechanisms in the early Universe, including the 
$SA$ co-annihilation (green), the Higgs resonance (blue), and $S S\to W W^*$ annihilation (red). 
We find that the loop correction to $m_{W}$ is dominated by the oblique parameter $T$ so that a large mass splitting between the charged Higgs bosons and neutral Higgs bosons can enhance $W$-boson mass sizably. Besides, we note that the mass ratios of $m_{S}/m_{H^\pm}$ and $m_{A}/m_{H^\pm}$ have to be less than one, i.e., $m_{S}/m_{H^\pm}<0.5$ and $0.35<m_{A}/m_{H^\pm}<1$, since they can produce a negative oblique parameter $S$ to further increase the $W$-boson mass. There is a clear gap at $m_{W}\sim 80.4\gev$ between gray and other three colors in both ($m_{S}/m_{H^\pm}$, $m_{W}$) 
and ($m_{A}/m_{H^\pm}$, $m_{W}$) planes,
owing to the fact that
the central value of $m_W$ from PDG2020 differs with the recent CDF II measurement
by $7\sigma$. Hence, they do not overlap in $m_{W}$-axis at the $95\%$ significance level. 

\begin{figure}
\includegraphics[width=8cm,height=8cm]{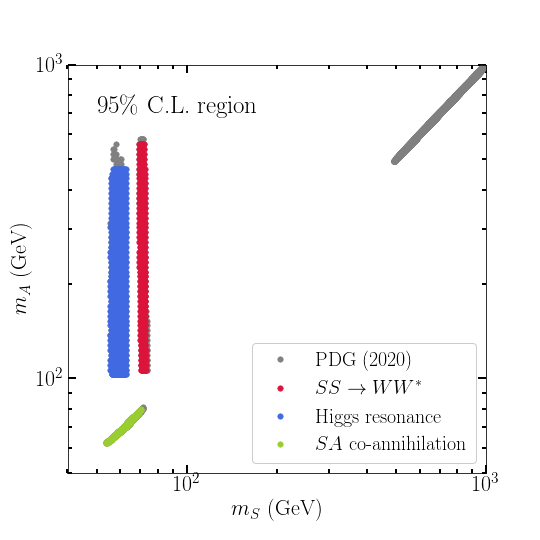}
\includegraphics[width=8cm,height=8cm]{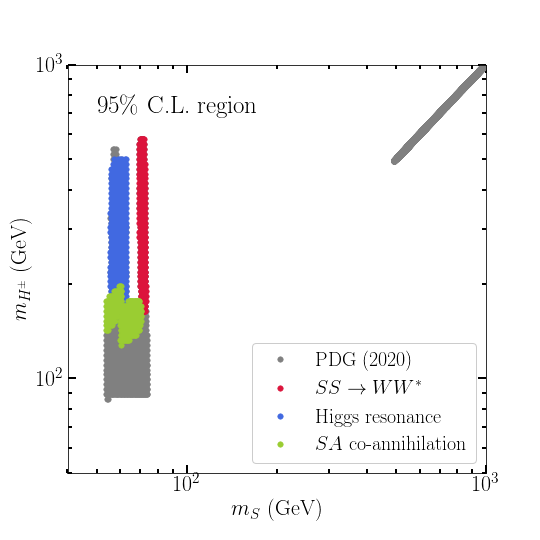}
\caption{Same as Fig.~\ref{fig:mass_ratio}, but on the planes of $m_{S}$ versus $m_{A}$, and $m_{S}$ versus $m_{H^\pm}$. 
Note that the low $m_{H^{\pm}}$ region (i.e., $\leq 120$ GeV) as well as the high $m_{S}$ region (i.e., $\geq 74$ GeV), 
despite of viable for the data of PDG2020, have been ruled out by $m_{W,{\rm CDF-II}}$.}
\label{fig:mass}
\end{figure}

In Fig.~\ref{fig:mass}, we present the allowed $95\%$ regions for above two cases on the planes of $m_{S}$ versus $m_{A}$ as well as $m_{S}$ versus $m_{H^\pm}$. It can be seen that the DM mass $m_S$ is bounded within the range of $54-74$ GeV. The previous higher mass region of $m_{S}>500\gev$ for $S-A-H^\pm$ co-annihilation is excluded. To account for the new CDF II $W$-boson mass, the mass differences between the charged Higgs bosons and neutral Higgs bosons should be enhanced; 
hence the mass degeneracy needed by the $S-A-H^\pm$ co-annihilation is broken.
The explicit correlations between the oblique parameters and new mass spectra can be found 
in Supplemental Material, where the relevant multileptons plus
missing energy signature~\cite{Belanger:2015kga,ATLAS:2019lff,Belyaev:2022wrn} are also discussed. 
Therefore, only three different favored parameter space for the DM relic density remains: 
$SA$ coannihilation with $m_{S}\approx m_{A}$ (green region), 
the Higgs resonance with $m_{S}\approx m_{h}/2$ (blue region), 
and the off-shell annihilation of $SS\to WW^*$ (red region). 
The first two mechanisms are in general with small couplings but  
the four-point interaction $\propto \lambda_3=2\left(m_{H^\pm}^2 - \mu_2^2\right)/v^2$
can efficiently govern DM annihilation for $m_{S}>60\gev$ even though one of $W$-bosons is off-shell. 
One can see that a kink of the coannihilation region, 
induced by the four-point interaction, appears at $m_{S}\sim 60\gev$ in Fig.~\ref{fig:mass}.

\begin{figure*}[htbp]
\begin{centering}
\subfloat[]{
\includegraphics[width=0.33\textwidth]{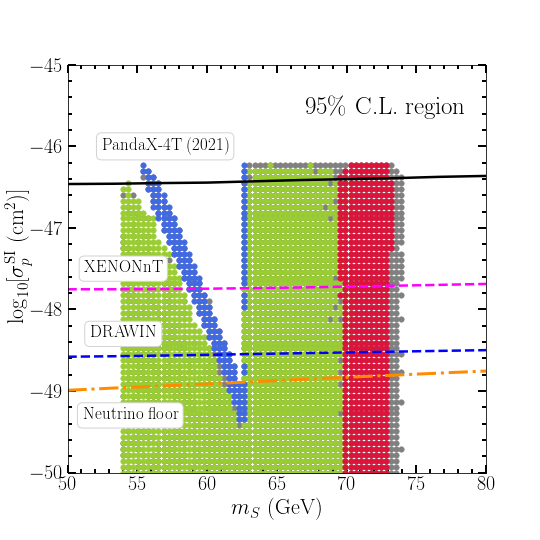}
\label{fig:prospect1}
}
\subfloat[]{
\includegraphics[width=0.33\textwidth]{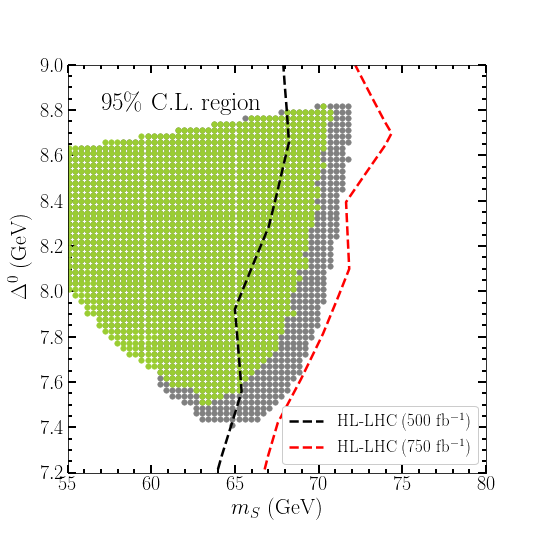}
\label{fig:prospect2}
}
\subfloat[]{
\includegraphics[width=0.33\textwidth]{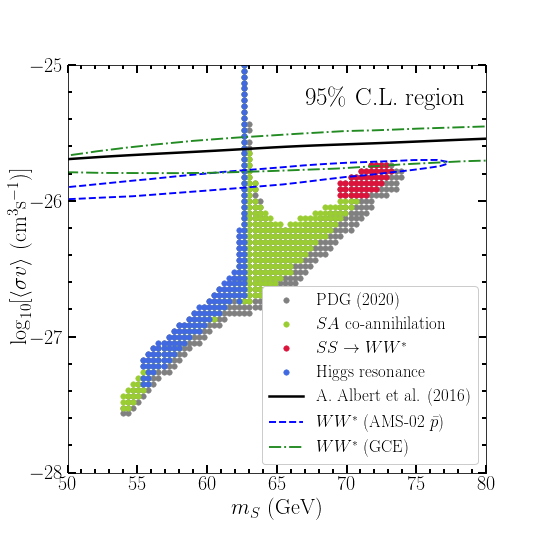}
\label{fig:prospect3}
}
\caption{The 95\% favored regions of 
$\sigsip$ [panel~(a)], $\Delta^0$ [panel~(b)], and $\sigmav$ [panel~(c)]
as a function of dark matter mass (i.e., $m_{\rm S}$). 
The Higgs resonance region and the $SA$ coannihilation region can be efficiently probed by 
the ongoing dark matter direct detection experiments~\cite{XENON:2020kmp,Schumann:2015cpa}
and the High Luminosity Large Hadron Collider~\cite{ATLAS:2019lng}, respectively. 
The $SS\rightarrow WW^*$ parameter region as well as a small fraction of the first and second regions 
are in agreement with that needed to reproduce the GeV $\gamma$-ray excess in the Galactic center and/or the GeV antiproton anomaly (the $2\sigma$ constraint counters are adopted from \cite{Zhu:2022tpr}).
}
\label{fig:prospect}
\end{centering}
\end{figure*}

In Fig.~\ref{fig:prospect}, we show the prospects of testing the above three favored scenarios in future collider and dark matter experiments. In the case of the Higgs resonance, we find that the DM-Higgs coupling is within the range of $|\lambda_S|<0.003$. The survived region covers the range of $4\times 10^{-50}~{\rm cm^{2}}\leq \sigma_{ p}^{SI}\leq 5\times 10^{-47}~{\rm cm^{2}}$, which has been tightly constrained by the latest PandaX-4T limit. Most of the remaining parameter space will be probed by the DM direct detection experiments in the near future. 
In contrary, a large portion of the survived parameter space of $SA$ coannihilation and $SS \rightarrow WW^*$ regions 
is below the so-called neutrino floor and hence beyond the scope of the conventional DM direct detection experiments. 
Fortunately, these two scenarios can be probed in future collider and DM indirect detection experiments. To be specific, due to a small mass splitting between $S$ and $A$ (see middle panel), we recast the current exclusion limit from the LHC search for the compressed supersymmetry~\cite{ATLAS:2019lng}, and also show the expected bound for the integrated luminosity $750$~fb$^{-1}$. We can see that the $SA$ coannihilation region can be fully covered at the future LHC. 
In addition, we note that the DM annihilation cross section can be around the thermal cross-section $\sigmav\sim 10^{-26}$~cm$^3 s^{-1}$ in $SA$ coannhilation, Higgs resonance and $SS\rightarrow WW^*$ regions. For the first two cases, the DM mass $m_S$ has to be in the range of $ 61~{\rm GeV} < m_S < 64$ GeV. On the other hand, the DM mass in $SS\rightarrow WW^*$ case lies in the range of $69~{\rm GeV} < m_S < 73$ GeV. Interestingly, such mass ranges coincide with the dark matter mass needed to account for the Galactic Center excess and Alpha Magnetic Spectrometer AMS-02 antiproton excess~\cite{Zhu:2022tpr}. Therefore, if all these intriguing anomalies are further confirmed, the i2HDM would be a natural model to accommodate them. 

\section{Conclusion}
%{\it Conclusion.}
The Collider Detector at Fermilab collaboration has just reported their latest measurement of $m_{W}$ with unprecedented accuracy, whose value is above the standard model prediction at a significance level of $7\sigma$. 
Such a $W$-boson mass anomaly, if confirmed by other experiments in the future, would point toward the presence of new physics that can be related to dark matter. 
The simplest dark matter model to account for the anomaly is the i2HDM. After addressing current theoretical and astrophysical constraints, we obtain viable regions of $m_{S}/m_{H^\pm}<0.5$ and $0.35<m_{A}/m_{H^\pm}<1$, for which the mass degeneracy of coannihilation at $m_{S}>500\gev$ has been broken. 
This is remarkably different from the pre-2022 data.
As a result, the heavy dark matter mass region has been thoroughly excluded and the charged Higgs must be heavier than $S$ and $A$. The DM mass is inferred to be between $54\gev$ and $74\gev$ and the thermal relic density was governed by the process of either the Higgs resonance, or $SA$ coannihilation, or $SS\to WW^*$ annihilation. The $m_{W,{\rm CDF-II}}$ favored dark matter mass range is well consistent with being a weakly interacting massive particle or WIMP, which is the most extensively discussed dark matter candidate. Encouragingly, the GeV $\gamma$-ray excess in the Galactic center and the possible GeV antiproton excess do consistently suggest a dark matter particle within the same mass range. Further dedicated efforts are highly needed to explore whether these two astrophysical signals and the $m_{W, \rm CDF-II}$ indeed have the common origin. Finally, it should be mentioned that, though the above conclusions are drawn with $m_{W, {\rm CDF-II}}$, the adaption of the latest full electroweak precision data~\cite{Lu:2022bgw} yields rather similar results, as shown in the Supplemental Material.

\section*{Acknowledgments}
%{\it Acknowledgments.}
We appreciate Andrew Fowlie for his kind and helpful discussions. This work was supported in part by the National Natural Science Foundation of China (No. 11921003 and No. U1738210) and by the Key Research Program of the Chinese Academy of Sciences (No. XDPB15).

\appendix

\section*{Supplemental Material}

\section{Additional information on the distribution of the physical parameters}
\label{app:supp}

Here we present more complete information on the distribution of the physical parameters constrained 
by current experimental/astrophysical constraints.

%%%%%%%%%%%%%%%%%%%%%%%%%   F   I   G   U   R   E   %%%%%%%%%%%%%%%%%%%%%%%%%%%%
\begin{figure}[ht!]
\centering
\includegraphics[width=8cm,height=8cm]{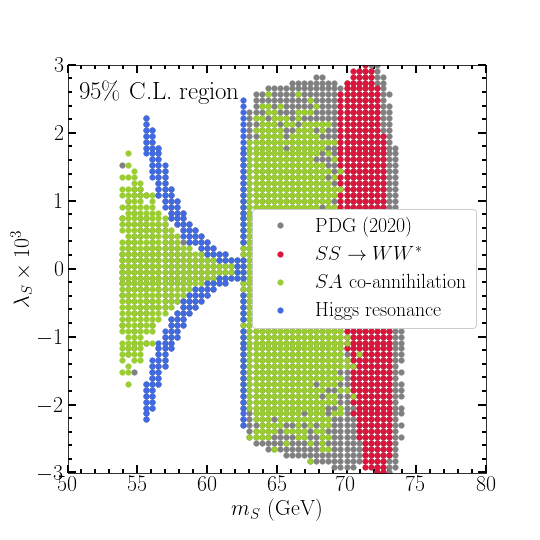}
\caption[]{The 95\% C.L. allowed regions: $SA$ coannihilation (green), the Higgs resonance (blue), and $S S\to W W^*$ annihilation (red) on the plane of $m_{S}$ versus $\lambda_S$. The gray region is for the data of PDG2020. Clearly, $\lambda_S$ is very small.
}
\label{fig:lambdaS}
\end{figure} 
%%%%%%%%%%%%%%%%%%%%%%%%%%%%%%%%%%%%%%%%%%%%%%%%%%%%%%%%%%%%%%%%%%%%%%%%%%%%%%%%

In Fig.~\ref{fig:lambdaS}, we show the DM-Higgs coupling as a function of $m_S$. 
The PLANCK relic density constraint results in a small favored region  
because other regions are corresponding to DM relic abundance $\Omega_\chi h^2$ 
overproduced. 
Because of the correlation $\Omega_\chi h^2 \propto 1/\sigmav$, a tiny coupling requires specific mechanisms, 
such as the $SA$ coannihilation (green), 
the Higgs resonance (blue) and $SS\to WW^*$ annihilation via four points interaction (red), 
to boost the annihilation cross section. 
The PDG2020 data has already imposed stringent constraints on $\lambda_{S}$ and the 
inclusion of $m_{W,{\rm CDF}}$ has just narrowed down the viable region slightly.

\begin{figure*}[htbp]
\begin{centering}

\subfloat[]{
\includegraphics[width=0.4\textwidth]{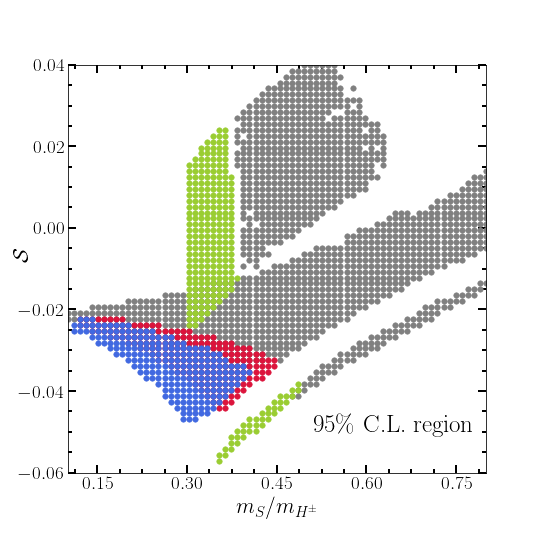}
\label{fig:stu3}
}
\subfloat[]{
\includegraphics[width=0.4\textwidth]{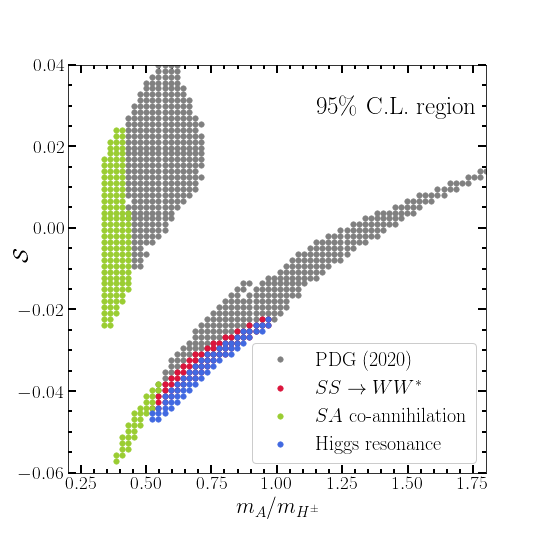}
\label{fig:stu1}
}\\
\subfloat[]{
\includegraphics[width=0.4\textwidth]{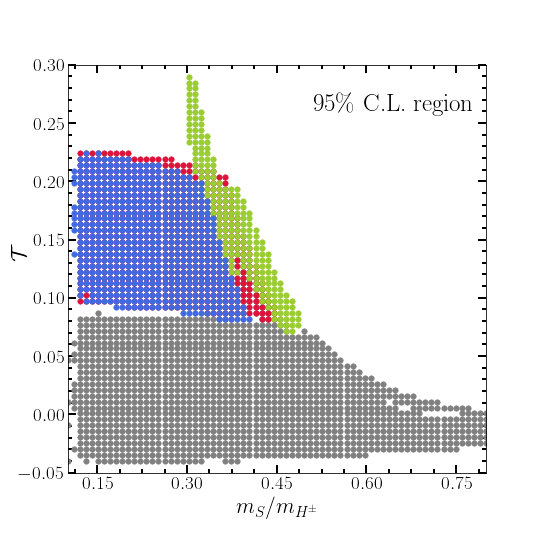}
\label{fig:stu4}
}
\subfloat[]{
\includegraphics[width=0.4\textwidth]{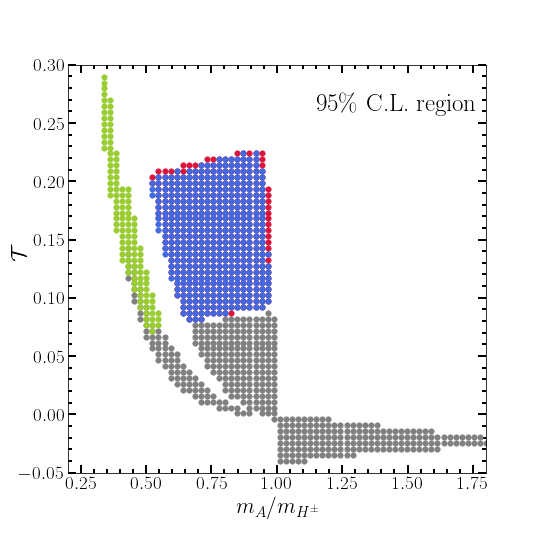}
\label{fig:stu2}
}
\caption{The correlations between oblique parameters and mass spectra. The CDF II $W$-boson mass measurement plays a key role in shaping the viable ranges of $\Sm$, $\Tm$ and $\Um$. To reproduce the data, $\Sm$ should be negative or almost zero if positive, while $\Tm$ is relatively high.}
\label{fig:STU}
\end{centering}
\end{figure*}

Since $\Delta m_W^{2}$ is proportional to  $-\Sm$, $\Tm$ and $\Um$ [see Eq.~(1) in the main text], the sizeable $W$-boson mass anomaly thus calls for a negative $\Sm$ or almost zero if positive and a relative large $\Tm$. 
Note that in i2HDM usually the absolute value of $\Um$ is smaller than that of $\Tm$ by one or two orders. 
In particular, a positive $\Sm$, which is only allowed in the $SA$ coannihilation scenario, would favor a high $\Tm$. 
The correlations between oblique parameters and mass spectra are displayed in Fig.\ref{fig:STU}. As anticipated, $m_{W,{\rm CDF-II}}$  sets stringent constraints on the ranges of $\Sm$ and $\Tm$ and we have $\Sm \leq 0.025$ and $\Tm \geq 0.07$.

%Clearly, the regions satisfying all current constraints are  $m_{\rm S}/m_{H^\pm}<0.5$ and $0.35<m_{\rm A}/m_{H^\pm}<1$ (see Fig.\ref{fig:STU}).  

\begin{figure*}[htbp]
\begin{centering}
%\subfloat[]{
%\includegraphics[width=0.4\textwidth]{plot_allmSmA.png}
%\label{fig:ms_ma_all}
%}
%\subfloat[]{
%\includegraphics[width=0.4\textwidth]{plot_allmSmHc.png}
%\label{fig:ms_mhc_all}
%}\\
\subfloat[]{
\includegraphics[width=0.4\textwidth]{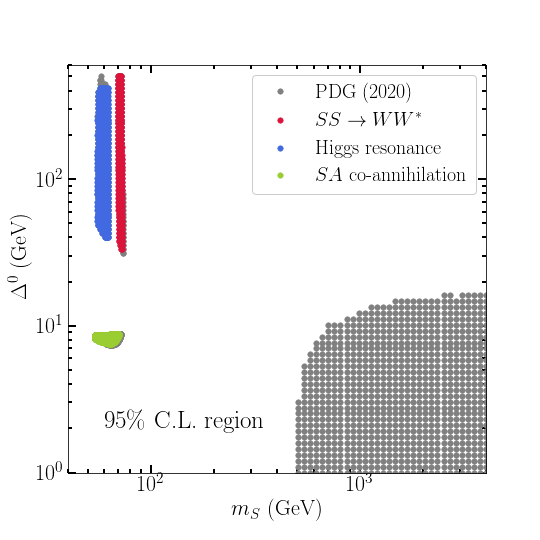}
\label{fig:delta0}
}
\subfloat[]{
\includegraphics[width=0.4\textwidth]{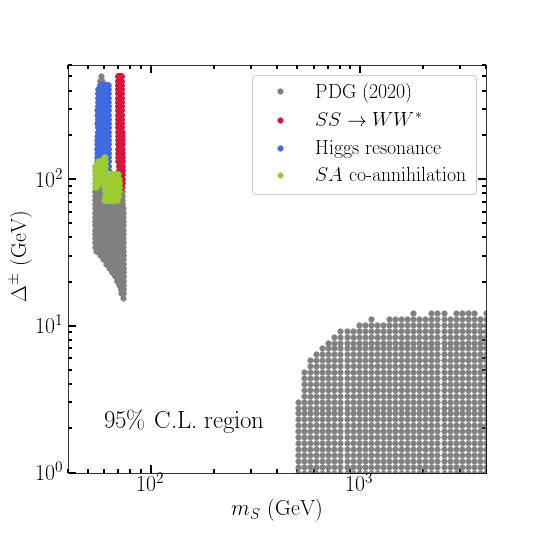}
\label{fig:deltapm}
}

\caption{The favored 95\% regions of $\Delta^{0}$ [panel~(a)] and $\Delta^{\pm}$ [panel~(b)]
as a function of the dark matter mass (i.e., $m_{\rm S}$). Different colors represent different physical processes governing the thermal relic density, as labeled in the plots.
The regions of $m_S>500\gev$ are also presented but they have been excluded by the CDF II $m_W$ measurement. 
}
\label{fig:full}
\end{centering}
\end{figure*}

In Fig.~\ref{fig:delta0} and Fig.~\ref{fig:deltapm}, 
we present the $\Delta^0$ and $\Delta^\pm$ as a function of $m_S$ in a wide range of $40\gev<m_S<4\tev$. 
In the absence of the CDF II $W$-boson mass measurement data, $m_S>500\gev$ is allowed under   
the coannihilation condition $m_S\approx m_A \approx m_{H^\pm}$, 
as found in the literature~\cite{Dolle:2009fn,Goudelis:2013uca}.  
Such a massive DM region, however, has been convincingly excluded by  $m_{W, \rm CDF-II}$. 
% as a comparison. 
For the $SA$ coannihilation scenario, we have a $\Delta^{0}$ clustering in the range of $7.5-8.8\gev$. 
The mass splitting $\Delta^{0}\sim 30-520$ GeV is needed for the other two scenarios. 
The CDF II $m_W$ anomaly also requires a $60\gev<\Delta^\pm<530\gev$.
Although a large $\Delta^\pm$ can enhance the coupling of $H^\pm$ pair production 
as searched in the multi-leptons plus missing energy signature~\cite{Belanger:2015kga,ATLAS:2019lff}, 
the heavy $H^\pm$ makes the cross section suppressed that the sensitivity
can only reach $\Delta^{\pm}\sim 
%\mathcal{O}
50 \gev $~\cite{Belyaev:2022wrn} at $m_S>50\gev$.

\section{Results with full electroweak precision data}
Since the new $m_W$ data from CDF II is slightly in tension with other electroweak precision measurements, 
it is interesting to check what happens with the combined analysis.  In this subsection, we therefore present the results based on the likelihood which includes all the electroweak precision measurements as done in PDG2020. 
For the full likelihood (EWPT2022),  
we adopt the new covariance matrix provided in Ref.~\cite{Lu:2022bgw} where the CDF II $m_W$ data is included in their global fitting. The updated oblique parameter are $\Sm=0.06\pm 0.1$, $\Tm=0.11\pm 0.12$ and 
$\Um=0.14\pm 0.09$, and the correlation coefficients of $(\Sm, \Tm)$, $(\Sm,\Um)$ and $(\Tm,\Um)$ 
are $0.90$, $-0.59$, and $-0.85$, respectively. 
We note that the full EWPT2022 likelihood sets slightly weaker constraints in comparison to that with the sole $m_W$ data. 
In addition, EWPT2022 likelihood enlarges our global minimum $\chi^2$ from $6.87$ to $14.01$.

 \begin{figure*}[htbp]
\begin{centering}
\subfloat[]{
\includegraphics[width=0.4\textwidth]{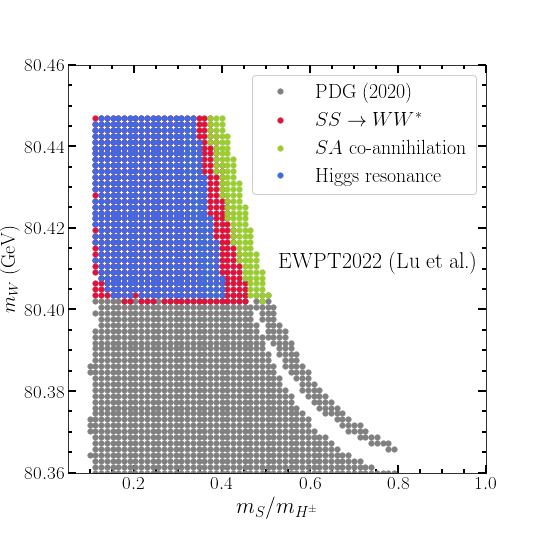}
\label{fig:ms_ma_all2}
}
\subfloat[]{
\includegraphics[width=0.4\textwidth]{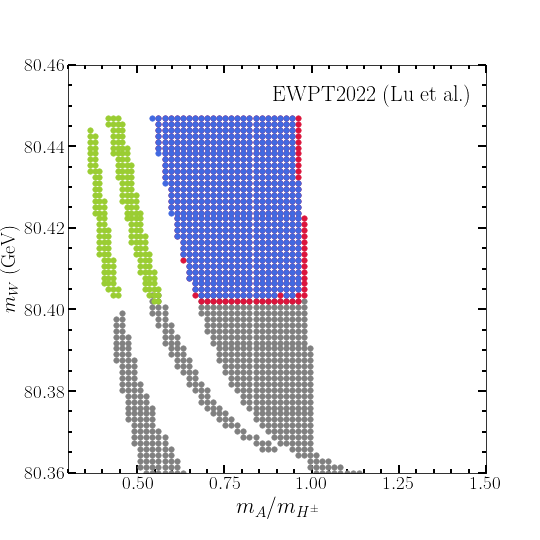}
\label{fig:ms_mhc_all2}
}\\
\subfloat[]{
\includegraphics[width=0.4\textwidth]{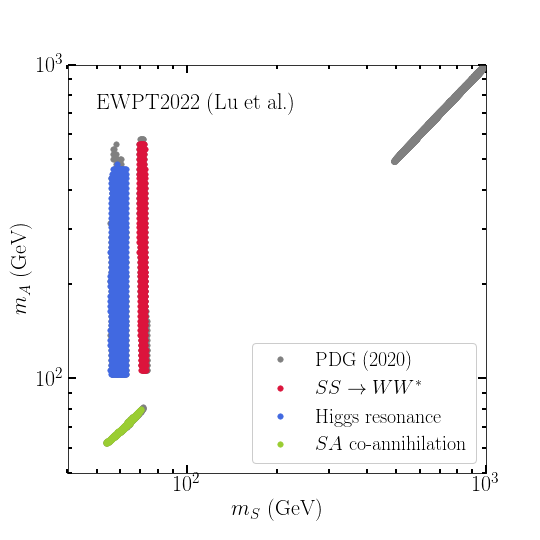}
\label{fig:delta02}
}
\subfloat[]{
\includegraphics[width=0.4\textwidth]{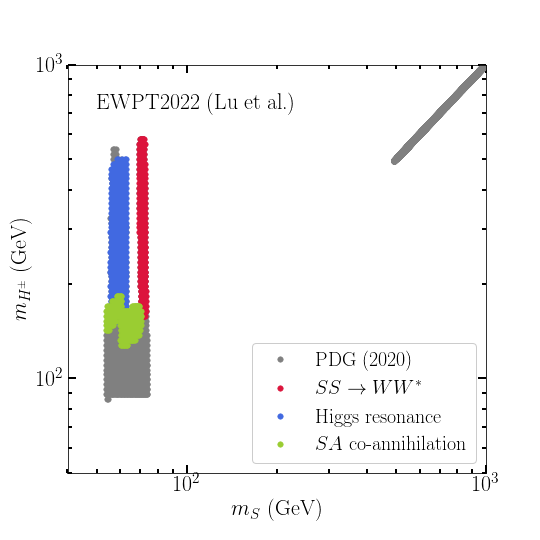}
\label{fig:deltapm2}
}

\caption{
The same as Fig.~1 and Fig. 2 in the main text but with the constraints of the new global EW fit~\cite{Lu:2022bgw}. 
}
\label{fig:full_vSTU}
\end{centering}
\end{figure*}
 
In Fig.~\ref{fig:full_vSTU}, we update the $2\sigma$ distributions shown in Fig.1 and Fig.2 with the constraints of EWPT2022.   
In comparison to Fig.1, now the $95\%$ allowed regions of $m_W$ 
have been (slightly) shifted downward
%Hence, comparing with Fig.~1 in the main text with Fig.~\ref{fig:ms_ma_all2} and \ref{fig:ms_mhc_all2}, 
and the gap between PDG2020 and CDF II results disappears. 
%In addition, the small portion of the allowed region located at the higher $m_W$ is also 
%gone once the full EWPT2022 likelihood is introduced. 
Anyhow, our other results and conclusions are essentially unchanged by comparing Fig.2 in the main text with Fig.~\ref{fig:delta02} and \ref{fig:deltapm2}.  The same holds for Fig.3 and for simplicity here we do not show the updated plots.

%%%%%%%%%%%%%%%%%%%%%%%%%%%%%%%%%%%%%%%%%%%%%%%%%%%%

\end{document}